\documentclass[a4paper]{jpconf}
\usepackage{graphicx}
\usepackage{lineno}
\begin{document}
\title{Optics robustness of the ATLAS Tile Calorimeter}

\author{R. Pedro on behalf of the ATLAS Collaboration}

\address{LIP, Av. Professor Gama Pinto 2, 1649-003 Lisboa Portugal}

\ead{rute.pedro@cern.ch}

\begin{abstract}
TileCal, the central hadronic calorimeter of the ATLAS detector is composed of plastic scintillators interleaved by steel plates, and wavelength shifting optical fibres. The optical properties of these components are known to suffer from natural ageing and degrade due to exposure to radiation. The calorimeter was designed for 10 years of LHC operating at the design luminosity of $10^{34}$~cm$^{-2}$s$^{-1}$. Irradiation tests of scintillators and fibres have shown that their light yield decrease by about 10\% for the maximum dose expected after 10 years of LHC operation.
The robustness of the TileCal optics components is evaluated using the calibration systems of the calorimeter: Cs-137 gamma source, laser light, and integrated photomultiplier signals of particles from proton-proton collisions. It is observed that the loss of light yield increases with exposure to radiation as expected. The decrease in the light yield during the years 2015-2017 corresponding to the LHC Run 2 will be reported.
The current LHC operation plan foresees a second high luminosity LHC (HL-LHC) phase extending the experiment lifetime for 10 years more. The results obtained in Run 2 indicate that following the light yield response of TileCal is an essential step for predicting the calorimeter performance in future runs. Preliminary studies attempt to extrapolate these measurements to the HL-LHC running conditions.
\end{abstract}

\section{Introduction}
\begin{figure}[t!]
\includegraphics[width=0.32\textwidth]{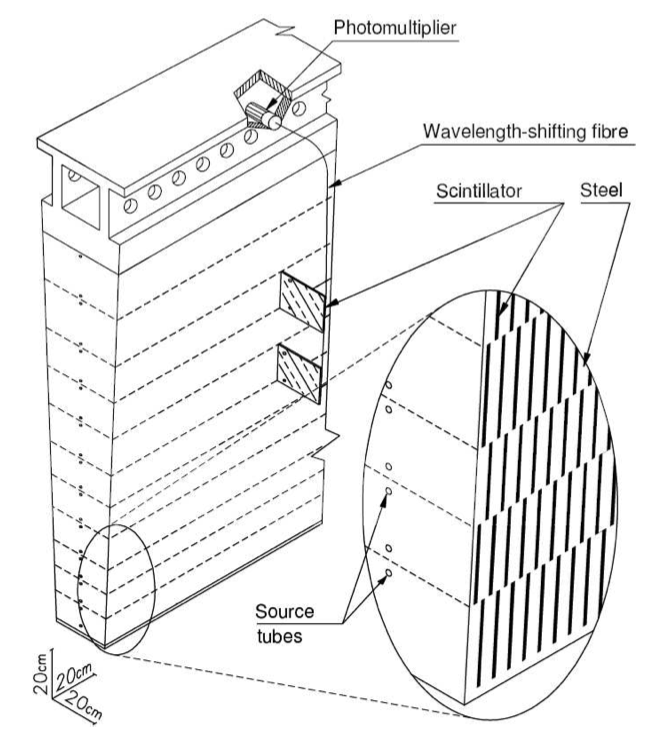}
\includegraphics[width=0.68\textwidth, trim=0 150 0 0, clip=true]{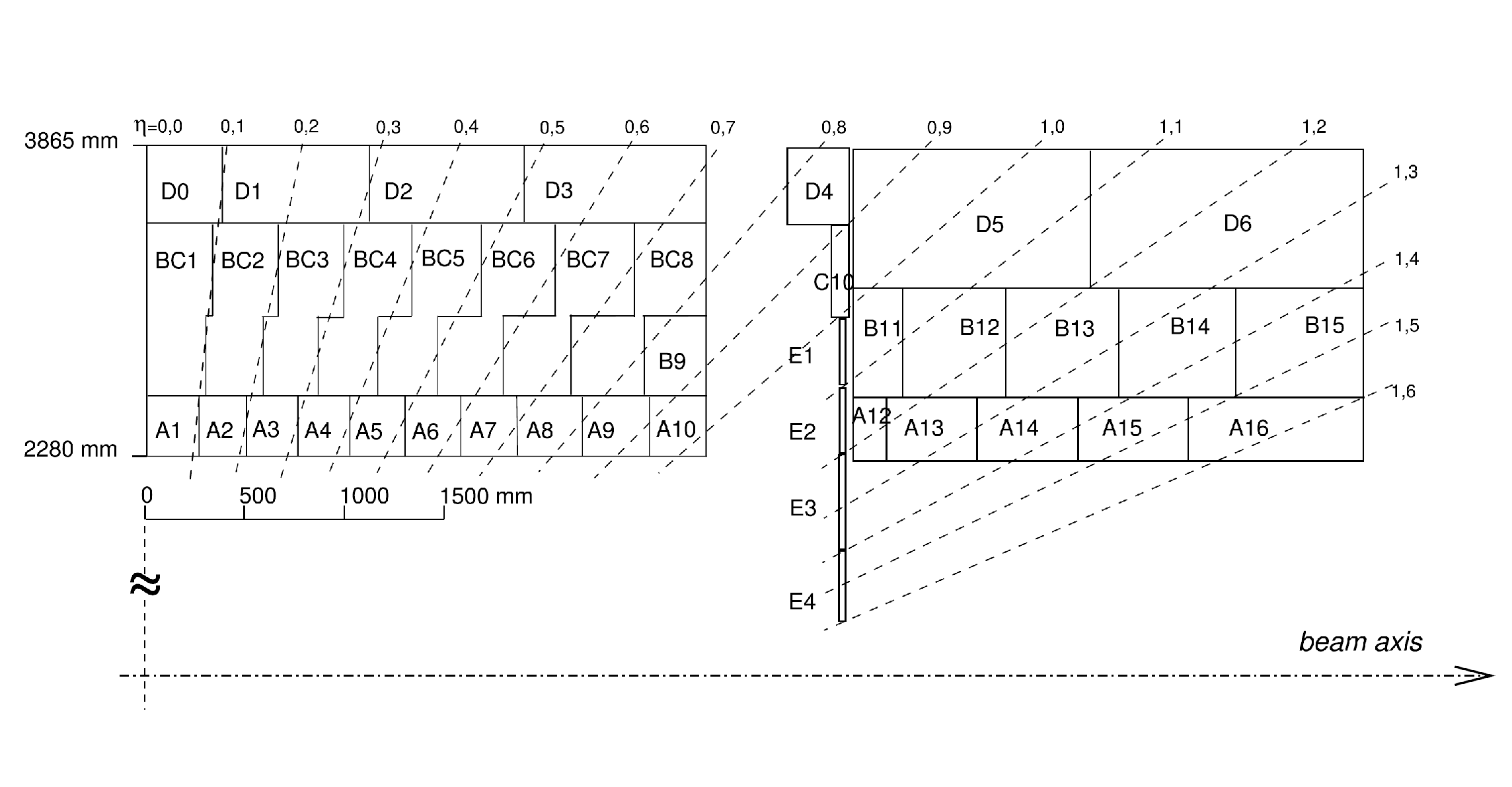}
\caption{\label{fig:TileSegm} Left: Schematic view of the optical readout of the Tile Calorimeter. The light produced by the plastic scintillator tiles is collected by  WLS optical fibres from two tile edges and guided to PMTs. Right: Segmentation in depth and $\eta$ of the Tile Calorimeter modules in the long barrels and in the extended barrels. Taken from~\cite{TileCalReadiness}.}
\end{figure}

The Tile Calorimeter of the ATLAS experiment is composed of 11 radial layers of steel absorbers interleaved with plastic scintillator tiles as active material (Fig.~\ref{fig:TileSegm})~\cite{ATLAS,TileTDR}. The light is collected by wavelength shifting (WLS) fibres from two edges of the tiles and guided to Hamamatsu R7877 photomultiplier tubes (PMTs). The tiles are 3~mm thick and have a trapezoidal shape of various sizes (10 to 19~cm height by 20 to 40~cm wide) and are polystyrene (PS)-based (PSM-115 or BASF165H) doped with 1.5\% PTP and 0.05\% POPOP. The WLS fibres (Kuraray Y11 MSJ) have 1~mm diameter and are also PS-based, and have a length of 80~cm to 2~m.

Each PMT reads out a bundle of fibres connected to the tiles composing a cell. TileCal is radially segmented into 3 cell layers: A, BC and D, ordered from the innermost layer to the outermost one (Fig.~\ref{fig:TileSegm}). Along the azimuthal direction, 64 modules split equally the $\phi$ space. The cell size across $\phi$ and $\eta$ is $0.1\times 0.1$ ($0.2$ in the D layer). Special scintillators called the E-cells, 6~mm thick, are placed in the gap/crack region of the detector. These are made of a different material than the tiles of the barrels: BC-408 (polyvinyl toluene-based) for E1 and E2 and UPS-923A (PS-based) for E3 and E4. 

Studying the optics robustness of TileCal is of crucial importance in view of the HL-LHC run since, in particular, scintillators and WLS fibres cannot be replaced, the only exception being the E-cells.

\section{Past studies of optics robustness}
The robustness of all optical components was studied in detail at the time of the detector design. The degradation in light yield of scintillators and WLS fibres due to natural ageing was determined to be below 1\%/year~\cite{ScintQual,FibreNatAgeing} using artificial accelerated ageing techniques. 

\begin{figure}[t!]
\includegraphics[width=0.5\textwidth, trim=0 17 0 0, clip=true]{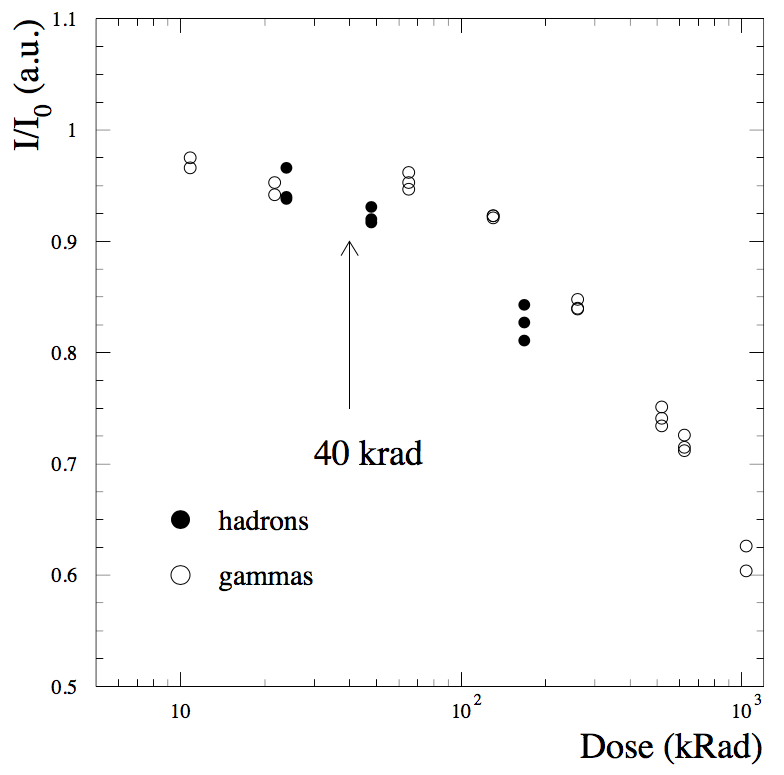}\hspace{2pc}%
\begin{minipage}[b]{0.45\textwidth}\caption{\label{fig:ScintillatorsIrradiations}Relative light yield of a scintillator one month after irradiation with gammas from Co-60 (6~rad/s) and secondary hadrons from interactions of 70~MeV proton beam with Al target (2 to 3~rad/s) as a function of the dose. The arrow at 40 krad indicates the dose expected at the end of 10 years of LHC operation at nominal conditions for the most exposed scintillators of TileCal. Taken from~\cite{ScintQual}.\vspace{1.5pc}}
\end{minipage}
\end{figure}

Scintillator tiles were irradiated with a Cobalt-60 gamma source and with secondary hadrons from the interactions of a 70~MeV proton beam with an Al target~\cite{ScintQual}. The dose rates of the irradiations were 60~mGy/s for gammas and 20 to 30~mGy/s for hadrons. The results (Fig.~\ref{fig:ScintillatorsIrradiations}) showed a decrease in the scintillation light yield of the tiles caused by the ionising radiation, and indicated a 10\% loss for 400~Gy, the expected dose after 10 years of LHC operation running at nominal conditions for cells placed at the innermost TileCal layer. Irradiations of scintillators coupled to WLS fibres were also done and the measured light yield reduction after 3~kGy was 19\%~\cite{TileTDR}. This value is of the same order of the one resulting from the irradiation of the scintillator alone shown on Fig.~\ref{fig:ScintillatorsIrradiations}, indicating that the scintillators are the most critical elements in the radiation hardness of the detector.

Exploring the TileCal calibration data is an opportunity to study the natural ageing and radiation damage to scintillators and WLS fibres in-situ conditions that, for instance, that in terms of dose rate levels are very difficult to reproduce in laboratory.

\section{Dose simulation}
\begin{figure}[t!]
\includegraphics[width=0.55\textwidth, trim=10 11 35 0, clip=true]{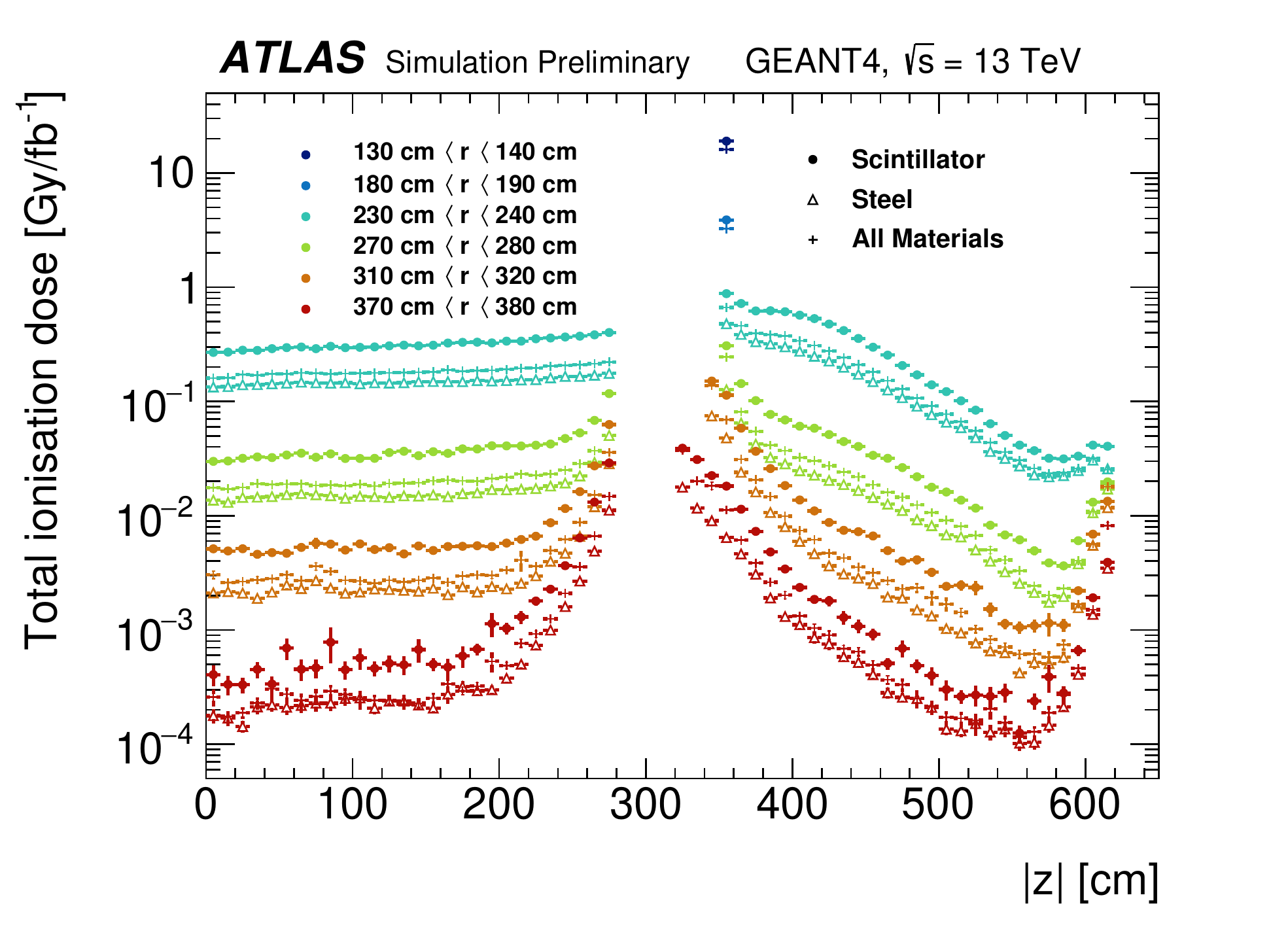}\hspace{2pc}%
\begin{minipage}[b]{0.40\textwidth}\caption{\label{fig:Tile_TID_prelim}Total ionisation doses from GEANT4 simulations of TileCal for pp collisions at $\sqrt{s}=13$~TeV for scintillating tiles, steel absorbers and all materials as average dose from the sum of individual doses. The simulation is based on 50000 inelastic pp events generated with PYTHIA 8 using the A3 tune and the NNPDF23LO PDF at $\sqrt{s}=13$~TeV normalised to a cross section of $\sigma_{\rm{inel}}$=78.42~mb and an integrated luminosity of L=1~fb$^{-1}$~\cite{DoseSimulation}.\vspace{0.5pc}}
\end{minipage}
\end{figure}

The total ionisation doses (TID) in the Tile Calorimeter were determined with GEANT4 simulation of proton-proton collisions at a centre-of-mass energy of $\sqrt{s}=13$~TeV (Fig.~\ref{fig:Tile_TID_prelim}). The dose is shown for different radius intervals, the larger TIDs correspond to the innermost radial layers, closer to the LHC interaction point, as expected. The dose is larger for the scintillator material than for steel due to the additional energy deposited by protons from the hydrogen atoms of the polystyrene scattered by neutrons.

The larger doses, up to 20~Gy/fb$^{-1}$, occur at $|z|\sim 360$~cm, where the E-cells are placed ($130<r<280$~cm), and in the smaller radial interval of the barrels ($230<r<240$~cm). The highly exposed region along $380<|z|<480$~cm corresponds to the A12, A13 and A14 cells of the extended barrel.

\section{TileCal calibration systems}
TileCal is equipped with dedicated calibration systems to monitor and calibrate each step of the readout chain~\cite{TileTDR}. A Cesium-137 source crosses the tiles, scanning the entire detector and allowing to determine the response of the entire readout chain. A laser system is used to perform the calibration and to monitor the PMTs and readout electronics and a charge injection system calibrates the digital readout. 

Besides the electronics that reads the fast signals from collisions, TileCal has an additional readout path with a RC integrator circuit. This path is used to read the signals generated by the Cs-137 scans. Signals from collisions are also integrated 20 to 25 times per run lumiblock and used to monitor the detector response. A lumiblock is a time interval for which the run had an approximate constant instantaneous luminosity, typically a minute. Events from minimum bias interactions, that dominate the pp collisions, lead to an average energy deposited in the calorimeter that is proportional to the instantaneous luminosity and so do their average integrated currents. Factorising this dependence out, for instance by normalising to the average integrated currents of a known stable cell, we can use collision data to monitor the detector response.

\section{PMT robustness}
\begin{figure}[!t]
\centering
\includegraphics[width=0.6\textwidth]{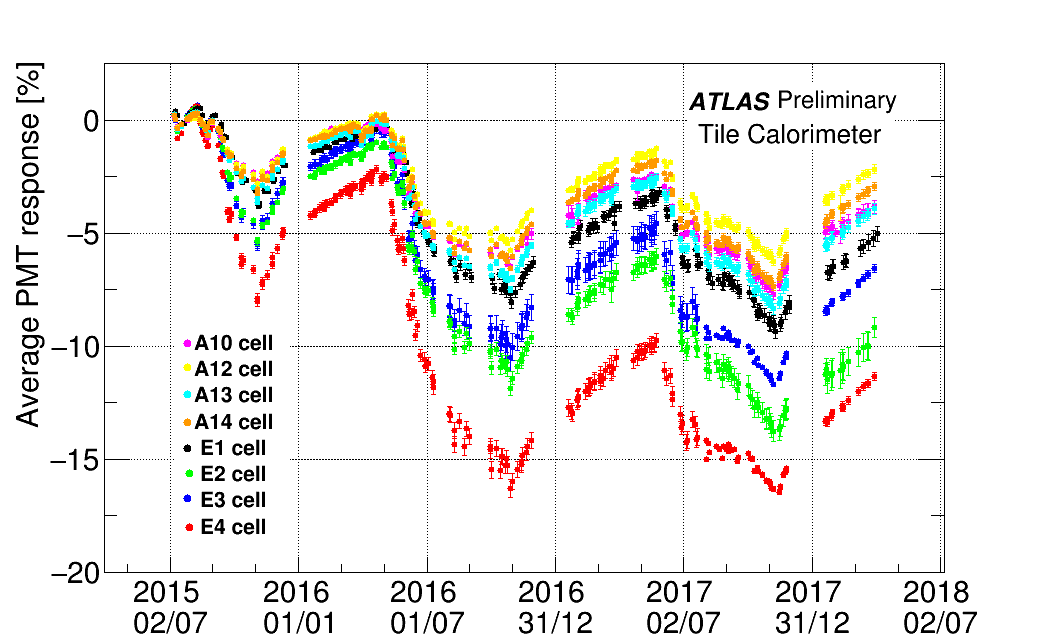}\hspace{2pc}%
\caption{\label{fig:PMT_evolution_2015_18}Variation of the average PMT response to the laser calibration pulses for PMTs reading the same type of cells as a function of time for the years 2015 to 2018. The response variation is computed with respect to an initial reference laser run taken on July 2015~\cite{TilePublicPage}.}
\end{figure}

The time evolution of the response of PMTs reading the cells most exposed to radiation, A10 to A14 and E-cells, was studied using data from the laser calibration system for the years 2015 to 2018 (Fig.~\ref{fig:PMT_evolution_2015_18}). The response evolution exhibits a couple of patterns. First, a down drift of the PMT response is observed during pp collision periods for all the cell types analysed. Secondly, this response degradation is partially recovered during LHC shutdowns. By fitting the data with a double exponential model featuring these two effects one concludes that the PMT response degrades at an approximately constant rate of 0.08\%/C on integrated charge. Taking for instance the A13 cell, for which the expected integrated charge at the end of the HL-LHC is extreme ($\sim$500~C) one expects at most around 40\% cumulated degradation of the PMT response.

\section{Radiation hardness of scintillators and WLS fibres}
\subsection{Monitoring with laser and Minimum bias integrated currents}
\begin{figure}[t!]
\centering
\includegraphics[width=0.65\textwidth]{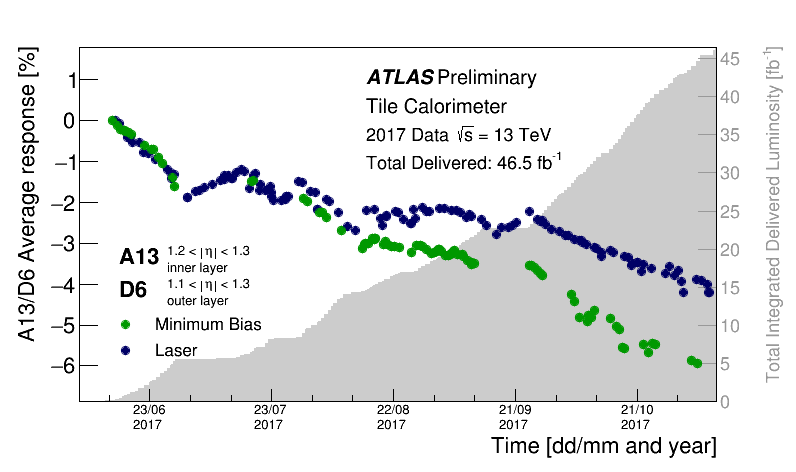}\hspace{2pc}%
\caption{\label{fig:2017_A13}Variation of the minimum bias integrated currents (green) and of the PMT response to the laser calibration pulses (blue) for 2017. The PMT and integrated currents are normalised to the D6 cell responses, stable within 1\%, and the data are averaged over the $64\times 2$ different modules in $\phi$. The response variation is computed with respect to an initial reference. Underlaid is the integrated luminosity as function of time (gray histogram)~\cite{TilePublicPage}.}
\end{figure}

The response of the entire optical chain of TileCal (scintillators, WLS fibres and PMTs) is assessed by monitoring the integrated currents from minimum bias events normalised to the D6 current to factorise out their linear dependence on the instantaneous luminosity. The reference cell D6 is found to be stable within 1\% during the time interval of the dataset analysed.
As discussed, the PMT response exhibits down-drifts matching the period of collisions and partial recovery during machine shutdowns. The response of scintillators, WLS fibres and PMTs altogether, follows the same trend but shows an increasing degradation with respect to the PMTs, Figure~\ref{fig:2017_A13}. At the end of 2017, the difference between the response to minimum bias events and to laser pulses is -3\% for the A13 cell. This is an effective way to monitor the response of scintillators and WLS fibres and to measure their light yield evolution.

\subsection{Relative light yield of scintillators and fibres}

The relative light yield $I/I_0$ of the scintillators and WLS fibres is computed from the difference in response to minimum bias events and laser pulses and accumulated along the years assuming no light yield evolution during LHC shutdowns. The $I_0$ is the measured light yield at the beginning of the period under analysis. Both measurements made with the laser system and with the integrator have a systematic uncertainty up to 1.2\%. These are propagated into the relative light yield uncertainty assuming no correlation between the two systematic sources, and summed quadratically.

\begin{figure}[t!]
\centering
\includegraphics[width=0.48\textwidth, trim=25 0 30 0, clip=true]{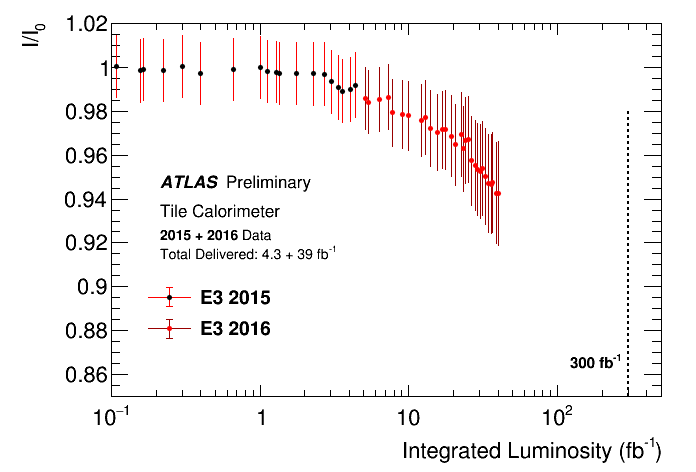}\hspace{1.5pc}%
\includegraphics[width=0.48\textwidth, trim=25 0 30 0, clip=true]{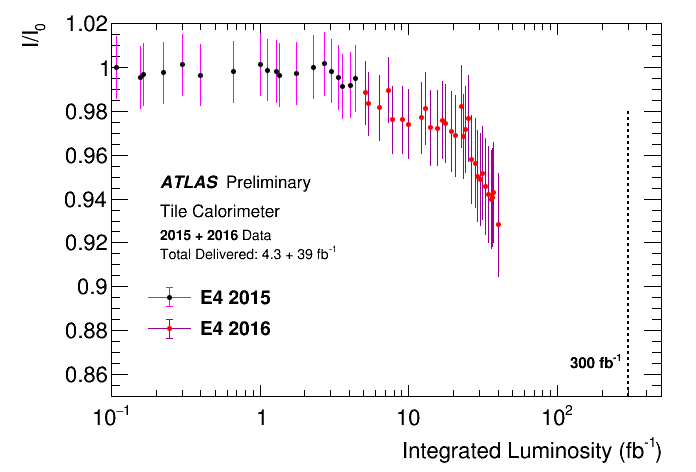}
\caption{\label{fig:LightYieldE3E4} The relative light yield $I/I_0$ of scintillators and wavelength-shifting fibres for the E3 (left) and E4 (right) cells as a function of the integrated luminosity for the year 2017. The vertical line points out the expected integrated luminosity in Run 3 only (300~fb$^{-1}$)~\cite{TilePublicPage}.}
\end{figure}

The $I/I_0$ of the E3 and E4 scintillators is plotted against the integrated luminosity for the years 2015 to 2016 (Fig.~\ref{fig:LightYieldE3E4}) and 8\% light loss is observed at the end of this period, where the total integrated luminosity was 43.3~fb$^{-1}$. These scintillators are made of a different material than the tiles composing the barrels and have also twice larger thickness. As stated, these aspects must be taken into account when establishing a direct comparison with the barrel cells. The E3 and E4 are the TileCal cells that face the most extreme irradiation conditions. They will be replaced during the next LHC long shutdown and are expected to be replaced at every long shutdown.

\begin{figure}[h!]
\centering
\includegraphics[width=0.48\textwidth, trim=25 0 30 0, clip=true]{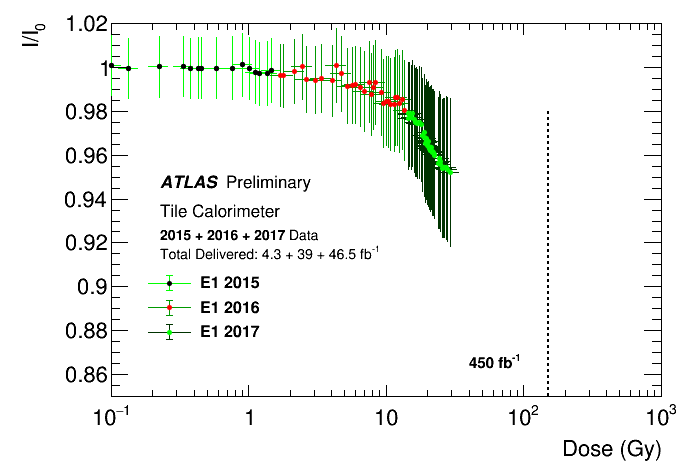}\hspace{1.5pc}%
\includegraphics[width=0.48\textwidth, trim=25 0 30 0, clip=true]{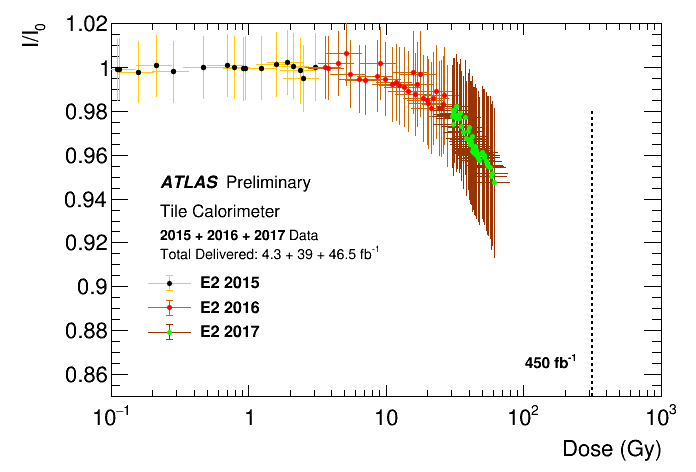}
\caption{\label{fig:LightYieldE1E2} The relative light yield $I/I_0$ of scintillators and WLS fibres for the E1 (left) and E2 (right) cells as a function of the dose for the years 2015 to 2017. The vertical line points out the expected dose by the end Run 3 (450~fb$^{-1}$ accumulated, accumulating with Run 1 and 2 luminosities)~\cite{TilePublicPage}.}
\end{figure}

\begin{figure}[t!]
\centering
\includegraphics[width=0.65\textwidth, trim=25 0 30 0, clip=true]{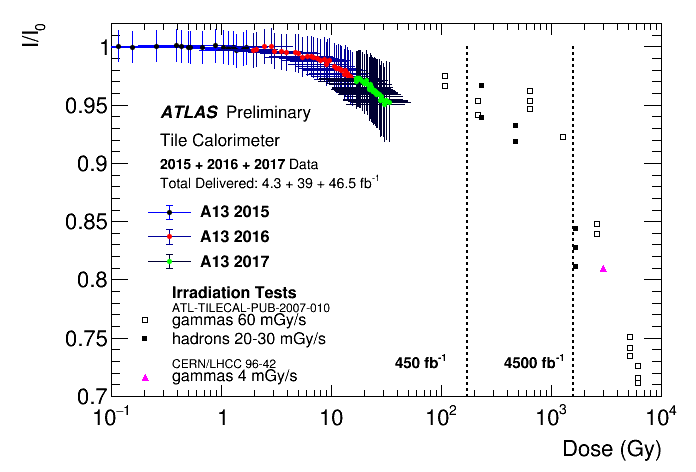}
\caption{\label{fig:LightYieldA13} The relative light yield $I/I_0$ of scintillators and wavelength-shifting fibres for the A13 cell as a function of the dose for the years 2015 to 2017. Data from scintillators irradiated with gammas (60~mGy/s) from Co-60 and with secondary hadrons (20-30~mGy/s) from the interaction of a 70~MeV proton beam with an
 Al target are also shown with black square markers. The pink triangle represents a measurement of the $I/I_0$ of a scintillator coupled to a WLS fibre after irradiation with a Co-60 gamma source (4~mGy/s)~\cite{TilePublicPage}.}
\end{figure}

The $I/I_0$ is shown as a function of the simulated dose in Fig.~\ref{fig:LightYieldE1E2} and~\ref{fig:LightYieldA13}. The dose simulation varies strongly along the detector and, given the large volume of the cells, the dose changes substantially within a cell. For the nominal dose the average value is taken and the RMS of the different dose values within the cell volume are represented as horizontal bars. No systematic uncertainty is assigned to the dose estimate. 
The loss of light yield increases with exposure to radiation, and the data points exhibit a well defined trend. 

Accumulating information from the years 2015 to 2017, one sees that the E1 and E2 scintillators and fibres loose 5\% of their light yield at the end of the period. These components of the detector are not expected to be replaced on the next LHC long shutdown and these data is useful to predict what will be the light output at the end of the Run 3, when the LHC is expected to have delivered 450~fb$^{-1}$. Based on this, a decision on the replacement of their PMTs can be taken, aiming at minimising the effect of the light yield loss. However, this extrapolation is limited by the large size of the error bars and this analysis will clearly benefit from the inclusion of the 2018 data, currently being taken.

Results for the A13 are compared with the results from irradiations made at laboratory of same material scintillators (square markers), and scintillators coupled to WLS fibres (triangle). The conditions of the laboratory irradiations are very far from the ones experienced by the scintillators mounted in the detector. Typical dose rates for cells in the A layer of TileCal are 1 to 10~mGy/h while they are $\mathcal{O}$(1 to 10~mGy/s) for the irradiations in the lab. Despite this, the light loss trend is similar. The in-situ data does not yet attain the dose region populated by the lab data, however some conclusions can be drawn already. Interpolating the laboratory data with the 1700~Gy estimated by up-to-date dose simulations for the A13 cell at the HL-LHC, one gets a lower limit of about 20\% on the light degradation in this cell. Moreover, TileCal data indicate a faster light loss rate than laboratory tests.

\subsection{Dose rate effects}

Dose rate effects on the radiation hardness of scintillator materials can be an important feature to consider given that CMS reports evidence for larger light output degradation with lower dose rate values~\cite{CMSdoseRateEffects,CMSphase2TDR}. It is however worth to note that the configuration of the tile-fibre coupling in CMS is very different from TileCal.  

\begin{figure}[t!]
\centering
\includegraphics[width=0.32\textwidth, trim=25 0 30 0, clip=true]{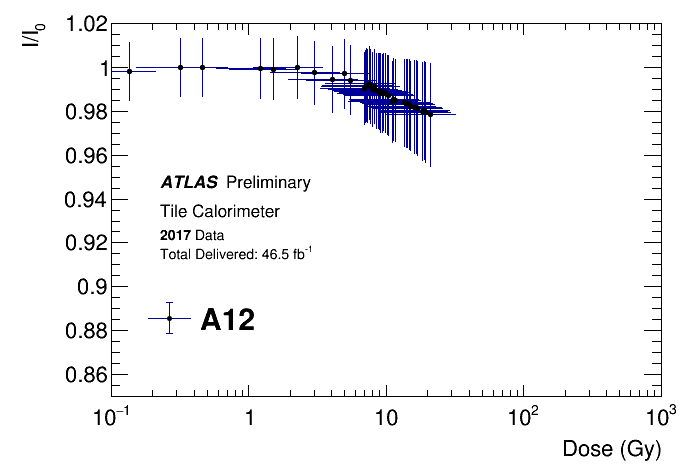}\hspace{0.5pc}%
\includegraphics[width=0.32\textwidth, trim=25 0 30 0, 
clip=true]{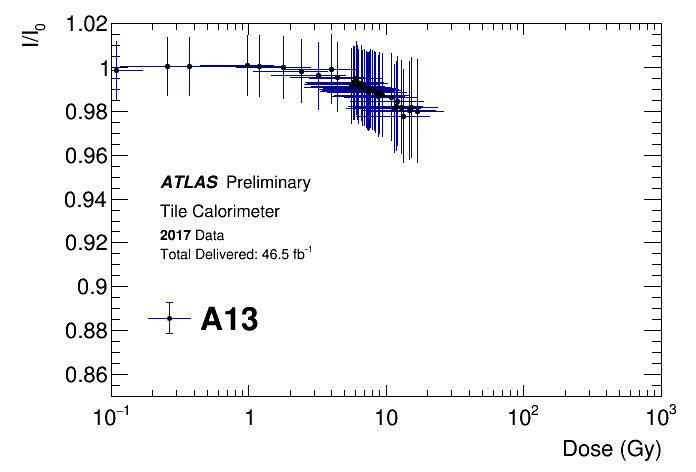}\hspace{0.5pc}%
\includegraphics[width=0.32\textwidth, trim=25 0 30 0, 
clip=true]{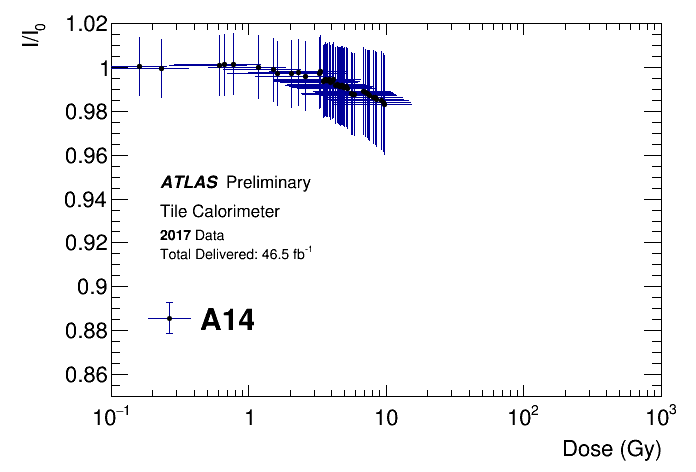}
\caption{\label{fig:DoseRateEffects} The relative light yield $I/I_0$ of scintillators and wavelength-shifting fibres for the A12 (left), A13 (middle) and A14 (right) cells as a function of the dose for the year 2017. The dose rate of the A12 cell is twice the dose rate of the A14 cell~\cite{TilePublicPage}.}
\end{figure}

Figure~\ref{fig:DoseRateEffects} shows a comparison of the relative light yield as a function of the dose for A12, A13 and A14 using data from 2017 only. This set of cells was chosen for having approximately same size scintillator tiles, read by about same sized WLS fibres (since the cells belong to the same layer, A), but with varying dose rate: the dose rate of A12 is twice the dose rate of A14, with A13 having an average value. The light loss is approximately the same for a common accumulated dose value: at 10~Gy, $I/I_0\sim0.985$. Although the sensitivity of the data is poor given the large uncertainties, one can observe that no huge dose rate effects are present, at this dose rate level.

\subsection{Expected performance at the end of HL-LHC}

The results presented so far focus on cells experiencing the most extreme irradiation conditions in TileCal. These cells were already subjected to a total dose that the least exposed cells, with much lower dose rate, will only attain at the HL-LHC phase. So, assuming that the dose rate does not play a substantial role on the radiation hardness model for the range of dose rates under consideration, the performance of the least irradiated cells can be inferred from the current data. The A13 cell is used as the reference because its material is common to the other cells in the barrel. The following is observed:

\begin{itemize}
\item Cells in the D layer are expected to have a dose between 4 and 200~Gy at the end of HL-LHC and a corresponding expected loss in light yield below 10\% for the majority of the cells. 
\item For the B/C layer the expected dose is 20 to 500~Gy and the expected degradation is below 15\% for most of the cells.
\item For cells in the A layer the expected dose is 300 to 1700~Gy; the reduction of systematics and more data are needed in order to make an extrapolation to this dose region.
\end{itemize}

\section{Conclusions}

The ATLAS TileCal calibration data from Run 2, 2015 to 2017, were analysed in view of determining the current performance of the optical components of the detector and estimating their robustness at the end of the HL-LHC phase.

Concerning photomultiplier tubes, a slow loss rate of 0.08\%/C on integrated charge is determined from the laser calibration data, and the expected degradation of the PMT response at the end of HL-LHC is at most $\sim$40\%, for the most stimulated photosensors.

Contrarily to photosensors, scintillator tiles and WLS fibres cannot be replaced and their expected degradation by the HL-LHC phase should be early determined. 

The accumulated degradation has a clear trend. This allowed to draw a model independent projection of the light yield loss of TileCal cells with the only assumption that dose rate effects are not substantial in the radiation hardness of organic materials. We conclude that 60\% of the TileCal cells (B/C and D layers) are expected to have a relative light yield of more than 85\% at the end of the HL-LHC phase. For the most exposed A-layer, the data are not enough to allow an extrapolation for such a far region. We need carefull understanding of the systematics to find ways to reduce their impact on the measurement, and more data in order to better define the fit model.

\section{Acknowledgements}
The author acknowledges the support of the grant CERN/FIS-PAR/0008/2017 from FCT, Portugal. The author thanks A. Henriques and A. Maio for the important contributions to this work.\\

\noindent Copyright 2018 CERN for the benefit of the ATLAS Collaboration. Reproduction of this article or parts of it is allowed as specified in the CC-BY-4.0 license.

\end{document}